\documentclass[%
 reprint,
 amsmath,amssymb,
 aps,
]{revtex4-1}

\usepackage{graphicx}
\usepackage{dcolumn}

\usepackage{multirow}
\usepackage{booktabs}
\usepackage{threeparttable}

\begin{document}

\preprint{APS/123-QED}

\title{Generating 10$\sim$40MeV high quality monoenergetic electron beams using a 5TW 60fs laser at Tsinghua University}
\thanks{Supported by National Natural Science Foundation of China (No. 11005063, No. 11375006 and No. 11175102) and Beijing Higher Education Young Elite Teacher Project}%

\author{J. F. Hua}
 \email{jfhua@tsinghua.edu.cn}
\author{L. X. Yan}
\author{C. H. Pai}
\author{C. J. Zhang}
\author{F. Li}
\author{Y. Wan}
\author{Y. P. Wu}
\author{X. L. Xu}
\author{Y. C. Du}
\author{W. H. Huang}
\author{H. B. Chen}
\author{C. X. Tang}
\author{W. Lu}
\affiliation{%
Department of Engineering Physics, Tsinghua University, Beijing 100084, China\\
}%


\date{February 14, 2014}

\begin{abstract}
A unique facility for laser plasma physics and advanced accelerator research has been built recently at Tsinghua Universtiy. This system is based on Tsinghua Thomson scattering X-ray source (TTX)\cite{Ref16, Ref5}, which combining an ultrafast TW laser with a synchronized 45MeV high brightness linac. In our recent laser wakefield acceleration experiments, we have obtained 10$\sim$40MeV high quality monoenergetic electron beams by running the laser at 5TW peak power. Under certain conditions, very low relative energy spread of a few percent can be achieved. Absolute charge calibration for three different scintillating screens has also been performed using the linac system.An article usually includes an abstract, a concise summary of the work
covered at length in the main body of the article.
\begin{description}
\item[keyword]
laser-plasma accelerator, monoenergetic, electron charge
\item[PACS numbers]
52.38.Kd, 52.59.Bi, 41.75.Jv
\end{description}
\end{abstract}

\pacs{Valid PACS appear here}
\maketitle





\section{Introduction}  
Plasma acceleration based on wakefield concept\cite{Ref17,Ref18} has made great strides in the past decade. By utilizing table-top 10TW-PW laser systems, high quality monoenergetic electron beams with energies up to few GeV\cite{Ref19, Ref6, Ref7, Ref8, Ref20, Ref9, Ref10} and energy spread of a few percent\cite{Ref3, Ref21, Ref4} have been obtained worldwide.
In this paper, we report our recent experimental results on laser plasma acceleration utilizing a newly built facility at Tsinghua University. This facility is based on Tsinghua Thomson scattering X-ray source (TTX), which combining an ultra-fast TW Ti:sapphire laser system with a synchronized 45MeV high brightness linac. By running the laser at 5TW 60fs, high quality 10$\sim$40MeV monoenergetic electron beams have been obtained. Under certain plasma conditions, very low relative energy spread of a few percent has also been achieved. The paper is organized as follows: in section II, the facility will be briefly overviewed; in section III, the experimental results will be presented in details: a summary is provided in section IV.
\section{Overview of Tsinghua laser plasma acceleration platform}
The laser plasma acceleration platform at Tsinghua University has three major components: the TW laser system, the 45MeV high brightness linac, and an experimental system combining a high vacuum chamber with a set of various optical and electric diagnostics. We next briefly describe these subsystems in details.

\subsection{TW laser system}
\begin{figure}[!htb]
\begin{center}
\includegraphics[width=8cm]{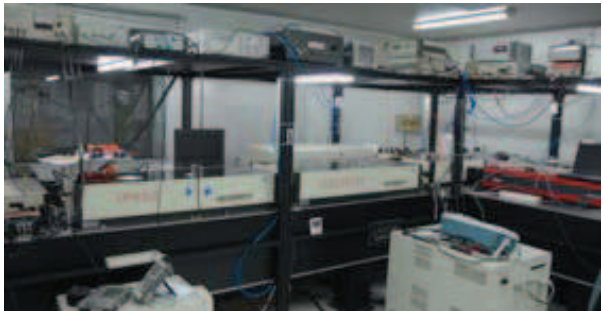}
\caption{ The ultrashort TW laser system.}
\label{figlaser}
\end{center}
\end{figure}

The TW laser system (Fig.~\ref{figlaser}), based on the standard CPA (Chirped Pulse Amplification) and  MOPA (Master Oscillator Power Amplifier) architecture, has six major components: an oscillator, a stretcher, a regenerative amplifier, two multi-pass amplifiers and a vacuum compressor~\cite{Ref15}.

The 20fs Ti:sapphire oscillator with a rep-rate of 79.3MHz is pumped by a frequency-doubled YVO$_4$ laser (Verdi-5). A fused-silica prism pair is used for intracavity dispersion compensation. The synchronization between the oscillator and the RF signal is achieved by a piezoelectric transducer (PZT) based feedback loop with timing jitter less than 200fs. Before entering the regenerative amplifier, the seed pulse goes through a Pockels cell pulse selector and then is stretched to about 300ps with a modified offener type stretcher. In the regenerative amplifier pumped by a frequency-doubled Q-switched Nd:YLF laser (Evolution-30), the laser pulse is amplified to about 3mJ with a rep-rate of 1kHz, and the bandwidth is reduced to 25nm. Another Pockels cell pulse cleaner is inserted before the first multi-pass amplifier to enhance the laser contrast against the ns prepulse and ASE. In the first four-pass amplifier pumped by a frequency-doubled Q-switched Nd: YAG lasers (Quanta-Ray Pro-350), the pulse is amplified to 100mJ in a bow-tie configuration. The laser beam is then split into two beams, with one beam (70mJ) as a seed for the final amplifier and another beam (30mJ) for third-order harmonic generation to drive the RF photocathode gun.
The final amplifier is another four-pass bow-tie amplifier pumped with two frequency-doubled Q-switched Nd:YAG lasers (Saga 230) from both ends. The amplification media is a 20mm-long, 20mm-diameter, normal-cut Ti:sapphire crystal with anti-reflective coating. With a combined pump energy of 2.8J, the laser pulse can be amplified to 1.0 J. A computer-controlled energy tuner made of a half-wave plate and a thin-film polarizer is utilized to adjust the laser pulse energy before compression. The laser pulse is expanded to 50mm-diameter beam size, and then is sent into the vacuum compressor. After compression, about 70\% of the laser energy can be sent into the experimental system. The final energy on target can reach up to 500mJ after taking into account the losses on transmission.

For laser-plasma interaction experiments, laser temporal contrast, focal spot quality and pulse duration are three crucial parameters. High temporal contrast is needed to avoid pre-plasma ionized by prepulses. High quality focal spot helps to excite the stable wake and to increase the energy coupling. Pulse duration should be compressed to as short as possible to increase the peak laser power. With the significant efforts in the past few years, our laser system has been thoroughly optimized for wakefield experiments. Typical parameters of our laser are listed in Tab.~\ref{laserpara}.

\begin{figure}[!htb]
\begin{center}
\centerline{\includegraphics[width=6cm]{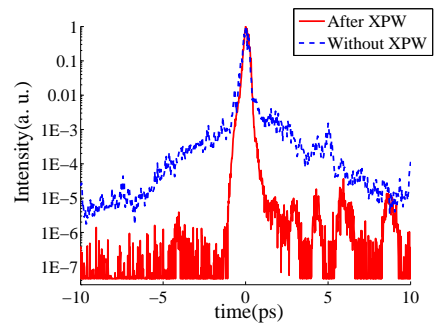}}
\caption{Contrast improvement by using XPW technology before other parameters are optimized. The ps contrasts are measured by a third-order autocorrelator with XPW generation (red line) and without XPW generation (blue dashed line).}
\label{contrast}
\end{center}
\end{figure}

The contrast at the nanosecond is measured by a fast photodiode coupled with ND filters, and the contrast at the picosecond is measured by a commercial third-order autocorrelator. The contrast has also been verified through experiments by observing pre-plasma using interferometry. At laser peak power, no pre-plasma has been found by scanning the delay time between the main laser pulse and the probe pulse. To further enhance the contrast, a technology called cross-polarized wave (XPW) generation has been tested, and a contrast enhancement of 2$\sim$3 orders at ps time has been confirmed, as shown in Fig.~\ref{contrast}.
\begin{table}[htbp]
\begin{center}
\caption{The parameters of TW laser system.}
\label{laserpara}
\footnotesize
\begin{threeparttable}
\begin{tabular*}{75mm}{cc}
\hline
\toprule Laser parameters & Value   \\
\hline 
{Central wavelength} & {800nm}\\     
{Pulse duration}& {60$\sim$70fs} \\   
{Reptition rate}& {10Hz} \\   
{Energy on target}& {500mJ} \\   
{Energy stability}& {1.5$\%$}\\
\multirow{3}{4cm}{Contrast ratio}& {$8.1\times10^6$@10ns}\\
 & {$1.9\times10^5$@10ps}  \\
 & {$2.5\times10^4$@1ps}  \\
\multirow{2}{4cm}{Focus spot (FWHM)\tnote{1} }& {16.6$\mu$m (Horizontal)}\\
 & {12.3$\mu$m (Vertical)}  \\
\multirow{2}{4cm}{Pointing stability (rms)}& {4.5$\mu$rad (Horizontal)}\\
 & {5.3$\mu$rad (Vertical)}  \\
 \hline
\bottomrule
\end{tabular*}
\begin{tablenotes}
\footnotesize
\item[1] Measured with an OAP mirror of 500mm focal length.
\end{tablenotes}
\end{threeparttable}
\end{center}
\end{table}
\begin{figure}[!htb]
\begin{center}
\centerline{\includegraphics[width=5cm]{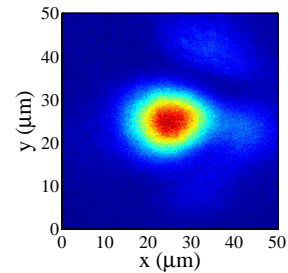}}
\caption{The horizontal and vertical focal sizes are measured to be 16.6$\mu$m and 12.3$\mu$m (FWHM) with f=500mm OAP mirror, respectively, and the 50\% energy is enclosed in the Gaussian-fitted focal spot. The RMS fluctuation of the focal spot is simultaneously measured to be 2.26$\mu$m (horizontal) and 2.65$\mu$m (vertical) respectively.}
\label{laserfocus}
\end{center}
\end{figure}
With an off-axis parabolic (OAP) mirror of 500mm focal length (II-VI Infrared), the laser beam has been focused to near-diffraction-limited size of 16.6$\mu$m (horizontal) and 12.3$\mu$m (vertical) FWHM, with 50\% energy enclosed in the Gaussian-fitted focal spot. The measurement is made by directly imaging the focal spot to a 10-bit CCD through a well corrected microscope objective lens. The RMS fluctuation of the focal spot has been measured to be 2.26$\mu$m (horizontal) and 2.65$\mu$m (vertical) respectively, corresponding to 4$\sim$5$\mu$rad angular pointing fluctuation.

Obtainable pulse duration of our laser system is about 60$\sim$70fs, due to the residual high-order dispersions within the laser system. An upgradation with a new 25fs front-end is planned later this year, and this may lead to much shorter pulse duration.
\subsection{Linear accelerator}
\begin{figure}[!htb]
\begin{center}
\centerline{\includegraphics[width=8cm]{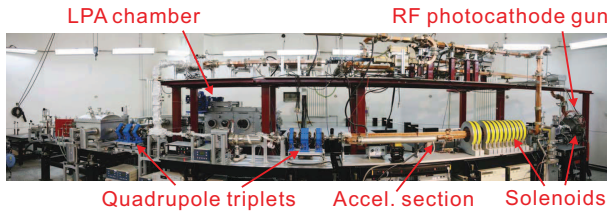}}
\caption{The 45MeV linac, which including a photocathode gun, solenoids, a 3m accelerating section, quadrupole triplets and a deflecting cavity.}
\label{linac}
\end{center}
\end{figure}

The 45MeV linac, as shown in Fig.~\ref{linac}, has two main components: a modified version of the BNL/KEK/SHI type 1.6 cell photocathode radio-frequency gun and a 3m SLAC-type traveling wave accelerating section. A UV laser pulse from the TW laser system irradiates the cathode nearly perpendicularly to generate an electron pulse. Solenoids are used for emittance compensation and beam envelope control. And the manipulation of beam can also be achieved by a pair of quadrupole triplets installed after the accelerating section. Utilizing the 20MW RF power, the maximum electron energy can reach to 45MeV with energy fluctuation of 1\% and energy spread of 1\%~\cite{Ref5}. Several beam diagnostics have been inserted into the beam line, including beam position monitors, high-resolution YAG screens, charge detectors (Faraday cups and ICT), an S-band RF deflecting cavity~\cite{Ref22} and an energy spectrometer.
\subsection{Experimental system}
\begin{figure}[!htb]
\begin{center}
\centerline{\includegraphics[width=6cm]{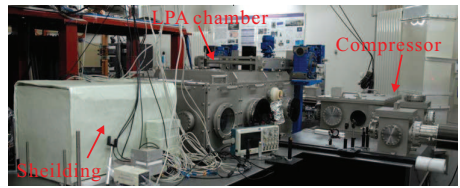}}
\caption{ The experiment station and the compressor in the experimental hall.}
\label{experimentalsystem}
\end{center}
\end{figure}

\begin{figure}[htbp]
\begin{center}
\centerline{\includegraphics[width=6cm]{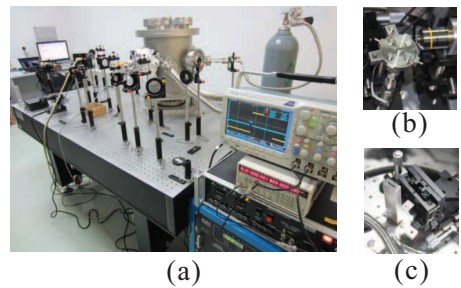}}
\caption{The development of plasma source. (a) The off-line gas density measurement platform; (b) A conical gas nozzle; (c) A 1mm$\times$10mm slit gas nozzle.}
\label{gasjet}
\end{center}
\end{figure}

\begin{figure}[!htb]
\begin{center}
\centerline{\includegraphics[width=6cm]{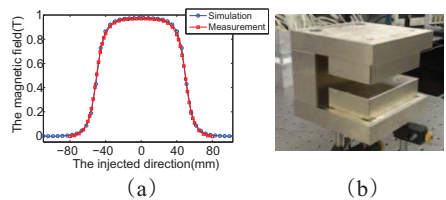}}
\caption{The compact permanent magnetic spectrometer. (a) The designed and measured magnetic field along the electron injection axis; (b) The magnetic energy spectrometer.}
\label{fig3}
\end{center}
\end{figure}

The experimental system have several main components, including a vacuum interaction chamber, radiation shielding, control and data acquisition (DAQ) and plasma sources. The vacuum interaction chamber has been properly designed to minimize the deformation during the vacuum pumping down process. Inside the chamber, all optical components sit on a reinforced optical breadboard which is locked down to an optical table under the chamber through bellows. Figure~\ref{experimentalsystem} shows the main components of the experiment station.
Various gas jet nozzles of Laval type have been designed and manufactured. Gas density profiles are measured by interferometry for Argon gas. Figure~\ref{gasjet} shows an off-line density measurement platform and two different gas nozzles.
A compact permanent magnetic spectrometer has been carefully designed to measure electron beams with a broad energy range and large momentum spread in typical laser plasma experiment. The magnet has a 15mm pole gap and has the magnetic field near 1T.
Figure~\ref{fig3} shows the simulated and measuremental data along the electron injection axis.
\section{Experimental results}
\begin{figure*}[!htb]
\begin{center}
\centerline{\includegraphics[width=16cm]{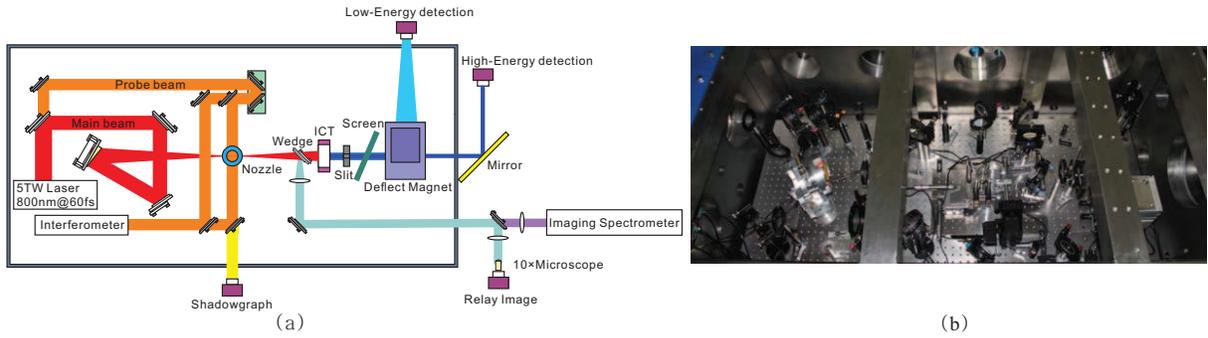}}
\caption{A schematic of our experimental arrangement (a) and the experimental setup inside the chamber (b). The main laser pulse is sent into the interaction chamber after compression, and then is focused close to the edge of gas jet nozzle by an OAP mirror. A probe beam, split off from the main beam, is used for interferometry and shadowgraph. The electron beam spectrum, the beam profile and the beam charge can be achieved by a compact magnetic spectrometer and the scintillating screens.}
\label{fig1}
\end{center}
\end{figure*}
\begin{figure}[!htb]
\begin{center}
\centerline{\includegraphics[width=7cm]{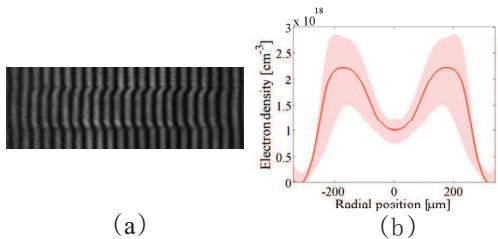}}
\caption{(a) The interferometer fringes at 3.3ns delay with a helium back pressure of 0.95Mpa; (b) The corresponding plasma density averaged over the longitudinal direction obtained by Abel inversion, which the shaded regions correspond to the standard deviation of the average over the interferogram in the longitudinal direction.}
\label{interferometer}
\end{center}
\end{figure}

Next we present our recent experimental results on laser wakefield acceleration.
Figure~\ref{fig1} shows a schematic of our experimental arrangement (a) and the actual setup inside the chamber (b). After compression, the main laser pulse is sent into the interaction chamber, and then is focused close to the edge of a 2mm-diameter helium gas jet nozzle by an OAP mirror with a focal length of 500mm. A Mach-Zehnder interferometer combined with a delay line is set up to measure the plasma electron density profile with a probe beam split off from the main beam. Figure~\ref{interferometer} shows a typical interferogram at 3.3ns delay (a) and the corresponding plasma density obtained by Abel inversion (b).

A scintillating screen (Mitsubishi: PI-200) is inserted into the beam path to diagnose the beam profile and the beam charge simultaneously. A 38$\mu$m-thick aluminum foil is placed before the screen to prevent the transmitted laser light.
A magnetic spectrometer can be moved in and out the beam path by a translation stage to measure the energy spectrum.
The spectrometer is equipped with two scintillating screens on two sides to monitor the low energy and the high energy range.
By running the laser at 5TW 60fs, we have obtained 10$\sim$40MeV high quality monoenergetic electron beams at a plasma density near $n_{e}=5\times 10^{19} cm^{-3}$. Details of the measurements of the energy spectrum, the electron beam profile and the beam charges are presented as follows.
\subsection{Electron energy spectrum and beam profile}
\begin{figure}[!htb]
\begin{center}
\centerline{\includegraphics[width=6cm]{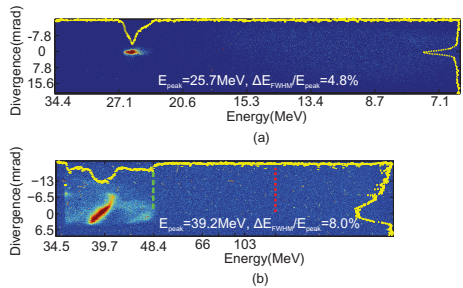}}
\caption{ Two examples of monoenergetic electron beams: (a) The peak energy of 25.7MeV with a FWHM relative energy spread of 4.8\%. (b) The peak energy of 39.2MeV with a FWHM relative energy spread of 8\%. The cut-off energy marked by the green dashed line is 49MeV, while the laser injection position is marked by the red dotted line.}
\label{fig4}
\end{center}
\end{figure}

\begin{figure}[!htb]
\begin{center}
\centerline{\includegraphics[width=6cm]{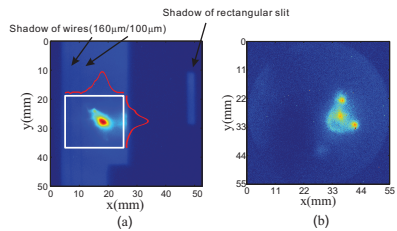}}
\caption{Electron beam profiles on the scintillating screens: (a) The FWHM beam size with 5.53mm (horizontal) and 4.12mm (vertical), contributing to the average divergence of 11mrad. The images of a rectangular slit and two metal wires (100$\mu$m and 160$\mu$m in diameter) are simultaneously produced by this accelerated beam. (b) Three different beamlets in a single shot.}
\label{fig2}
\end{center}
\end{figure}

In the energy spectrum measurement, a 3mm-wide rectangular tungsten slit is introduced into the beam path right before the spectrometer, which gives an uncertainty on the measured energy less than 0.2MeV for a 25MeV electron beam. The typical electron beam energy obtained in our experiment is in the range of 10$\sim$40MeV. Here we show two examples. In Fig.~\ref{fig4}(a), the electron beam has a peak energy of 25.7MeV and a FWHM relative energy spread of 4.8\%. The vertical divergence of the beam is 2.3mrad FWHM.
In Fig.~\ref{fig4}(b), the electron beam has a peak energy of 39.2MeV and a FWHM relative energy spread of 8\%. The vertical divergence of the beam is 10.3mrad FWHM. The cut-off energy of this beam is 49MeV, as marked by the green dashed line.
A typical electron beam profile is shown in Fig.~\ref{fig2}(a). This profile is obtained by a PI-200 screen placed 431mm downstream the gas jet, with a tilted angle of (61$\pm$2)$^\circ$ from the beam axis.
We also put a few metal wires with different diameters right before the screen to acquire high-resolution electron radiography.
Occasionally, multiple electron beams can be generated simultaneously in a single shot, and an example with three different beamlets is shown in Fig.~\ref{fig2}(b).

\subsection{Charge calibration and beam charge measurement}
To measure the electron beam charge accurately, we adopted a widely accepted method based on sensitive scintillating screens.
Several authors provide calibration data on different scintillating screens~\cite{Ref1, Ref2} using electron beams from linac.
They found two types of screens (PI-200 and Drz-high) are more sensitive than the Lanex regular screen. However directly adopting their calibration data in our experiment may cause some ambiguities because the exact setups and the properties of the CCD cameras are different from experiment to experiment. To avoid these ambiguities, we performed a similar calibration procedure using our in-house linac system. A relative shorter exposure time (100$\mu$s) is used in our calibration to reduce the background noise.

\begin{figure}[!htb]
\begin{center}
\centerline{\includegraphics[width=6cm]{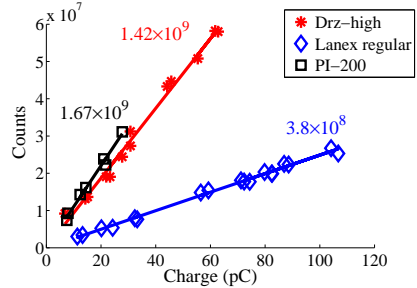}}
\caption{Absolute charge calibration of three different screens (red for Drz-high, blue for Lanex regular and black for PI-200) using electron beams from the linac. The makers show the measuremental data, and the gradients of the best fit lines are also provided.}
\label{fig5}
\end{center}
\end{figure}

The electron beam energy of the linac is fixed at 30MeV with a bunch duration of about 1ps, and an ICT (Bergoz ICT-055-070-05:1) is used to benchmark the electron charge. The electron charge is kept below the saturation threshold of the scintillating screens\cite{Ref1}. A cooled 16-bit CCD camera (Apogee: Alta U2) with a Nikon lens (focal length 60mm and F number 1/2.8) is used to record the signal, and a band pass filter at 546 nm is also applied to reduce the background light. In Fig.~\ref{fig5}, we plot the calibration data for three different screens. We found that the PI-200 and Drz-high screens are about three times more sensitive than the Lanex regular screen. The measured sensitivity of PI-200 is 1.67$\times10^9$ counts/sr/pC, while the sensitivity of Drz-high is 1.42$\times10^9$ counts/sr/pC. Based on our calibration, the charge for the electron beam shown in Fig.~\ref{fig2}(a) is 10.6pC, and the charges for energy measurements are around 0.2pC (shown in Fig.\ref{fig4}(a) and (b)).

\section{Conclusions}
A unique facility for laser plasma physics and advanced accelerator research has been built recently at Tsinghua Universtiy. This system is based on Tsinghua Thomson scattering X-ray source (TTX), which combining an ultrafast TW laser with a synchronized 45MeV high brightness linac. In our recent laser wakefield acceleration experiments, we have obtained 10$\sim$40MeV high quality monoenergetic electron beams by running the laser at 5TW peak power. Under certain conditions, very low relative energy spreads of a few percent can be achieved, which closes to the best published results~\cite{Ref3, Ref21, Ref4}. Absolute charge calibration for three different scintillating screens has also been performed using our linac system, and the typical charge of accelerated electron beams is in the picocoulomb range.
\vspace{10mm}
\acknowledgments{The authors would like to thank Dr. Z. Y. Wei, Dr. H. Teng, Dr. Z. H. Wang and Dr. J. L. Ma from Institute of Physics (CAS) for their great support on the laser system construction, Dr. J. Wang, Dr. H. -H Chu and Dr. S. -Y. Chen from National Central Universtiy (Taiwan) for their valuable information on the laser system optimization. We also thank to Dr. J. Zhang from China Institute of Atomic Energy for the use of their contrast measurement equipment.}
\vspace{10mm}

\bibliographystyle{unsrt}

\end{document}